\documentstyle[preprint,aps,floats,epsf,epsfig]{revtex}

\begin{document} 
\def\beq{\begin{equation}}
\def\eeq{\end{equation}}
\def\ov{\overline}
\thispagestyle{empty}
\draft
\markright{Revised \today}

\title{Large {  $\rm {\bf {\rm N_c}}$} Limit of 
Spin-Flavor Breaking \\ in Excited Baryon Levels}
\author{J. L. Goity}
\address{
Department of Physics, Hampton University, Hampton, VA 23668, USA \\
and \\
Thomas Jefferson National Accelerator Facility \\
12000 Jefferson Avenue, Newport News, VA 23606, USA.}

\maketitle  
\begin{abstract}
 Spin-flavor symmetry breaking in the levels of excited   Baryons are
 studied to leading order in the  1/${\rm N_c}$ expansion. This breaking
 occurs at zeroth order. For   non-strange Baryons with a single quark
   excited, it is shown that     
 to   first order of perturbation theory  the breaking
 is given by one 1-body operator (spin-orbit), and three 2-body operators, 
 all involving the orbital angular momentum of the excited quark.
 Higher-body operators can be reduced to that set of operators.
 As illustration, p-wave Baryons are briefly discussed.
\end{abstract}

\newpage

In the large ${\rm N_c}$   limit, 
QCD admits an  expansion in powers of $1/{\rm N_c}$. 
This is the only small  expansion parameter of the gauge theory 
identifiable in the non-perturbative  regime.
The coupling constant $\alpha_s$ scales 
  as $\alpha_0/{\rm N_c}$,
 where  $\alpha_0$ is  of order unity.
  In general, the coefficients of the 
  expansion are non-perturbative in $\alpha_0$
and, therefore, difficult to calculate. Despite this general difficulty, the contributions
 to different orders in the expansion can be systematically identified as 
 `t Hooft showed long ago$\cite{t'Hooft}$. The usefulness of the  
$1/{\rm N_c}$ expansion has been shown in the study of mesons in QCD in 
$1+1$ dimensions$\cite{1+1}$
where exact leading order solutions   have been obtained  
$\cite{1+1,Callan}$. 
Similar expansions have been shown to be very successful
  in different field theory models$\cite{Coleman}$. 
In QCD in 3+1 dimensions no definite progress has been made in implementing
 the expansion at the quantitative level. However, some  important                                                                                                                                                                                                                                                                                                                                                                                                                                                                                                                                                                                                                                                                                                                                                                                                                                                                                                                                                      
                                             
qualitative
 predictions can be drawn by  identifying the leading scaling 
in $1/{\rm N_c}$ of various   observables, for instance,
 $F_\pi\sim\sqrt{{\rm N_c}}$,
 $M_{{\rm Meson}}={\cal{O}}(1)$,  
  $M_{{\rm Baryon}}= {\cal{O}}( {\rm N_c})$,  
 excited meson widths   
 $\Gamma_{\rm Meson}={\cal{O}}(1/{\rm N_c})$, excited Baryon widths  
 $\Gamma_{\rm Baryon}={\cal{O}}( 1)$, OZI rule,
  validity of the valence quark picture up to 
  ${\cal{O}}(1/{\rm N_c})$ corrections, 
  classification of ${\cal{O}}(p^4)$ effective coupling constants in ChPT
   according to their large ${\rm N_c}$  behaviour, etc.

 Many of the interesting
questions in the large ${\rm N_c}$ limit 
can be found in the sector of Baryons, a sector of particular  
interest  due to its complexity.
 Baryons in the  large ${\rm N_c}$
limit consist of ${\rm N_c}$ valence quarks in the singlet representation 
of SU(${\rm N_c}$). 
It was shown by Witten$\cite {Witten1}$ that Baryons in 
the ground state are described in terms of a constituent quark picture
which can be implemented in the Hartree approximation neglecting spin-flavor dependent
interactions.
 Moreover, these ground state Baryons,  which  are in the totally symmetric representation 
of the spin-flavor group SU(2${\rm N_f}$)
(${\rm N_f}$:  number of flavors), can be described as Skyrme solitons as well
$\cite{Skyrme,Witten2}$. Both pictures turn out to be   equivalent$\cite {Manohar1}$.
Always in the sector of ground state Baryons, important relations have been 
derived for couplings of   pions$\cite {Sakita,Dashen1}$,
SU(3) breaking in the Baryon masses$\cite {Jenkins}$, and magnetic moments
$\cite{Lebed}$.

In this paper we   address a less explored sector, the excited Baryon sector.
 There are important questions one may ask concerning the spectroscopy and 
 the different transitions. 
Some   have   been  
 considered, like the question of strong decays of p-wave Baryons$\cite {Carone}$, 
and the  Adler-Weisberger sum rule in large ${\rm N_c}$
$\cite{Masperi}$.  Here we   study the leading corrections to the 
spectrum of non-strange Baryons, and the aim is to explore 
the spin-flavor violating pieces of the Baryon masses. 
The leading ${\cal{O}}({\rm N_c})$ 
piece of the masses is spin-flavor symmetric, and the leading 
corrections which break this 
symmetry are  of  
${\cal{O}}(1)$.   We  only address   Baryons where a single quark is excited, 
 with the rest ${\rm N_c}-1$ quarks
 left in the ground state. In the limit 
where the symmetry breaking 
can be taken as a perturbation, we   show that there are five 
different types of effective interactions  
that can give leading order spin-isospin symmetry breaking, namely, 
 spin-orbit   of Thomas precession type, 
    spin-isospin tensor, and three   operators involving spin, orbit and 
    isospin components. 
   Baryons with more than one excited quark or the case of three flavors 
 can be studied along similar lines.

\section{The Picture of Excited Baryons in Large ${\rm N_c}$}

In large  ${\rm N_c}$,  Baryons can be pictured as a color singlet  state
of ${\rm N_c}$ valence (constituent) quarks, with a subleading admixture of sea quarks.
 The wave functions are  required to be totally symmetric under the simultaneous 
interchange
of coordinate and spin-flavor indices due to color antisymmetrization.
The ground state Baryons form a tower of states described by a self-consistent Hartree 
wave function$\cite{Witten1}$:

\begin{equation}
\Phi(x_1,\;\xi_1;...;x_{{\rm N_c}},\;\xi_{{\rm N_c}})=
\prod_{i=1}^{{\rm N_c}} \psi(x_i)\;\chi_{_S}(\xi_1,...,\;\xi_{{\rm N_c}})
\end{equation}
where $\psi(x_i)$ are $\ell=0$  wave functions, $x_i$ 
the position of the $i^{th}$ quark and $\xi_i$ its spin-flavor quantum numbers. 
$\chi_{_S}(\xi_1,...,\;\xi_{{\rm N_c}})$ is the totally symmetric  tensor  of 
rank ${\rm N_c}$ in spin-flavor space. For two flavors,   
the ground 
state Baryons form a tower   
satisfying  the $I=J$ rule $\cite{Mattis}$.
A typical state in the tower can be expressed as follows$\cite{Luty}$:
\begin{equation}
\mid B\rangle=\frac{1}{{\rm N_c}!}\;\int\prod_{i=1}^{{\rm N_c}}\frac{d^3k_i}{(2\pi)^3}\; 
\tilde{\Phi}_B(k_1,\;\xi_1;...;k_{{\rm N_c}},\;\xi_{{\rm N_c}})\epsilon_{\alpha_1,...,\;
\alpha_{{\rm N_c}}} \;\hat{a}^{\dagger\;\alpha_1}_{\xi_1}(k_1)...
\hat{a}^{\dagger\;\alpha_{{\rm N_c}}}_{\xi_{{\rm N_c}}}(k_{\rm N_c})
\mid 0\rangle
\end{equation}
where $\hat{a}^{(\dagger)}$ are effective creation and annihilation operators of valence quarks, 
and $\alpha_i$ are color indices.

States with excitation energy of zeroth order in $1/{\rm N_c}$  correspond to having 
one or more quarks in an excited state$\cite {Witten1}$,
 while a large number of order ${\rm N_c}$ remain in the ground 
 state (\lq\lq core").  
In large ${\rm N_c}$  the core quarks are  described by the same wave functions
that describe the quarks in the ground state tower with the replacement of ${\rm N_c}$ by 
${\rm N_c}-n$, $n$ being the number of excited quarks. The classification of excited states in terms 
of spin-flavor $ SU(2{\rm N_f})$ is very convenient, as this group is a dynamical symmetry 
group of the ground state sector.
 We   consider  states where 
only a single 
quark is excited. Such states are contained in the totally symmetric representation,
 corresponding to the  $[{\rm N_c}]$ Young tableaux, and the mixed symmetry representation
corresponding to  the $[{\rm N_c}-1,1]$ tableaux. As mentioned above, for two flavors the first
 representation contains
   the $I=J$ states with unit multiplicity, while the latter contains $I=J$  and $ I=J\pm 1$ states
also with unit multiplicity. In terms of  the K-spin utilized in the induced representation 
method$\cite{Sakita,Cook,Dashen2}$, the first representation corresponds to K=0 and the latter to K=1.
The states in these representations are respectively given in terms of the 
wave functions:

\begin{eqnarray}
\Phi_S(x_1,\;\xi_1;...;x_{{\rm N_c}},\;\xi_{{\rm N_c}})&=&
\frac{1}{2\sqrt{N_c}}\sum_{j }
(\prod_{i\neq j}  \psi(x_i))\,\phi(x_{j})
\;\chi_{_S}(\xi_1,...,\;\xi_{{\rm N_c}})\nonumber\\
\Phi_{M}(x_1,\;\xi_1;...;x_{{\rm N_c}},\;\xi_{{\rm N_c}})&=&
\frac{1}{\sqrt{{\rm N_c} ({\rm N_c}-1)}}
\sum_{i\neq j} (\prod_{k\neq i,j}  \psi(x_k))\,
\psi(x_i)\phi(x_j)-i\leftrightarrow j )\nonumber\\
&\times& {\chi_{_M}}(\xi_i,\;\xi_j;\;\xi_1,...,\xi_{i-1},
\xi_{i+1},...,\xi_{j-1},\xi_{j+1},...,\xi_{{\rm N_c}})
\end{eqnarray}
where ${\chi_{_M}}$ is the irreducible rank ${\rm N_c}$ mixed symmetry tensor   
with the $[{\rm N_c}-1,1]$ tableaux, and $\phi(x)$ is the wave function of
 the excited quark.
It is easy to show that the Hartree equations for the two different spin-flavor
representations differ by exchange terms of    ${\cal{0}}(1) $. 
The description of states with a larger number of excited 
quarks is by obvious generalization.

Although it does not play a crucial role in our analysis,
the wave functions must be eigenfunctions  of the total momentum.
 The following projection
solves the CM motion problem present in the above wave functions:
\begin{equation}
\Phi(x_1,\;\xi_1;...;x_{{\rm N_c}},\;\xi_{{\rm N_c}})\rightarrow 
\int \Phi(x_1-\bar{x},\;\xi_1;...;x_{{\rm N_c}}-\bar{x},\;\xi_{{\rm N_c}})\;
\exp(-i P.\bar{x})\;d^3\bar{x}
\end{equation}

In this work we are interested in calculating matrix elements 
of spin-flavor breaking Hamiltonians.
Each Hamiltonian will be represented at the level of the effective theory by a 
sum of composite effective operators  of different dimensions expressed in terms of the 
effective quark  creation and annihilation operators and carrying
   the same relevant quantum numbers as the original QCD operator. These composite
 operators are classified in the $1/{\rm N_c}$ expansion according to 
their leading contributions to   matrix elements. 
Operators which contain $n$ creation and $n$ annihilation effective quark operators are 
called $n$-body operators. Since in order to build an $n$-body operator one has 
to exchange
at least $n-1$ gluons between quark lines,  each $n$-body operator has a coefficient
proportional to $\alpha_s^{n-1} = {\cal O} ({\rm N_c}^{-n+1})$. 
 The matrix elements of the $n$-body operator will
in general bring back some factors of $ {\rm N_c}$ which partially or totally eliminate 
that suppression.  Static operators have been   studied
in the case of ground state Baryons$\cite {Dashen2,Luty}$, where
 it is possible to establish
reduction rules   relating matrix elements of higher-body operators to
lower-body operators,  enormously simplifying the problem of obtaining the most general
form of the matrix elements of a given QCD operator in the effective theory. 
A  similar reduction applies to the static operators relevant to our analysis;
  higher-than-two-body
operators can in fact be reduced  to  
 one- and two-body operators as shown later.

One-body color singlet   static operators can be expressed as follows:
\begin{equation}
\hat{Q}^{(1)}_\Gamma=\int\frac{d^3k}{(2\pi)^3}\;
 \Gamma(\xi_1,\;\xi_2; \; k)\; \hat{a}^{\dagger\;\alpha}_{\xi_2}(-k) \, \hat{a}_{\alpha}^{\xi_1}(k),
\end{equation}
where $\Gamma$ is the kernel characterizing the operator. 
The matrix elements between   excited states $\mid B\rangle$ and 
$\mid B^{\prime}\rangle$ are then given by
 (assume the total momentum of the Baryon vanishes):
\begin{eqnarray}
\langle B^{\prime}\mid \hat{Q}^{(1)}_\Gamma \mid B\rangle&=&{\rm N_c}  
\int \prod_i\frac{d^3k_i}{(2 \pi)^3} \;  \Gamma (\xi_1,\;\xi_1^{\prime}; \; k_1)
\delta^3(k_1+...+ k_{N_c})\nonumber\\
&\times &
\tilde{\Phi}_{B^{\prime}}^{\ast}(k_1,\;\xi_1^{\prime};k_2,\;
\xi_2;...;k_{{\rm N_c}},\;\xi_{{\rm N_c}})
\;\tilde{\Phi}_{B}(k_1,\;\xi_1;k_2,\;\xi_2;...;k_{{\rm N_c}},\;\xi_{{\rm N_c}})
\end{eqnarray}

Similarly, two-body effective operators have the general form:
\begin{equation}
\hat{Q}^{(2)}_\Gamma= \int\prod_{i=1}^{4}\frac{d^3k_i}{(2\pi)^3}\; 
\Gamma(k_1,\;\xi_1;...;\;k_4,\;\xi_4) 
\;\hat{a}^{\dagger\,\alpha_2}_{\xi_4} (k_4) 
\,\hat{a}^{\dagger\,\alpha_1}_{\xi_3}(k_3)
\,\hat{a}_{\alpha_2}^{\xi_2}(k_2)
\,\hat{a}_{\alpha_1}^{\xi_1}(k_1)~~,
\end{equation}
and their matrix elements are given by:
\begin{eqnarray}
&&\langle B^{\prime}\mid \hat{Q}^{(2)}_\Gamma \mid B\rangle={\rm N_c}({\rm N_c}-1) 
  \int\prod_{i=1}^{{\rm N_c}-2}\frac{d^3k_i}{(2 \pi)^3}
\prod_{l=1}^{4}\frac{d^3q_l}{(2 \pi)^3}\;
 \Gamma(q_1,\eta_1;...;q_4,\eta_4)\;\delta^3(q_1+q_2-q_3-q_4)\nonumber\\ 
&\times& \tilde{\Phi}_{B^{\prime}}^{\ast}(k_1,\;\xi_1;...;\;k_{{\rm N_c}-2},
\;\xi_{{\rm N_c}-2}; q_3,\;\eta_3; q_4,\;\eta_4)
  \;  \tilde{\Phi}_{B}(k_1,\;\xi_1;...;\;k_{{\rm N_c}-2},\;\xi_{{\rm N_c}-2};
 q_1,\;\eta_1; q_2,\;\eta_2).
\end{eqnarray}
It is not difficult to extend   this to higher-body operators.

 \section{Leading Spin-Flavor Symmetry Breaking}

When spin-flavor dependent interactions are neglected, the excited 
Baryon spectrum shows degenerate spin-flavor towers. 
The  masses of the  towers are, as mentioned before, 
of  order   $ {\rm N_c}$.
Degeneracy is lifted at zeroth order in   $1/{\rm N_c}$ by  the effective Hamiltonian operators 
which we  identify and  
study in this section. Of course, in order to lift the degeneracy 
we need operators which are
non-trivial under spin-  and/or flavor-transformations.
Also, we are assuming here that the breaking of spin-flavor symmetry can be treated 
perturbatively, and therefore, we are only interested in taking matrix elements
between states which are degenerate in the symmetry limit. This in particular means
that   
one needs only   consider matrix elements between states with the same orbital 
angular momentum $\ell$ of the excited quark; 
this eliminates some operators  which are to be included  
 beyond first order perturbations. 
 
 In the following we  restrict  the discussion to two flavors.  
States in the different towers can be expressed more conveniently in the  
 basis
$\{\mid I_c;\;I^{T}\;I^{T}_3;\;\;S^{T} \; S^{T}_3,\;\ell\;m \rangle\}$, 
where $I_c$ and
$S_c$ are the isospin and spin quantum numbers  of the core,  
$I^{T}$ and $S^{T}$ the total isospin and spin, and $\ell$ is the 
orbital angular momentum. States with $I^{T}=S^{T}\pm 1$ belong to  
$K=1$ towers, while for $I^{T}=S^{T}$ states one has for     states
with $I^T<< {\rm N_c}/2$:
\begin{eqnarray}
\mid K=0, I^{T}=S^{T}, I^{T}_3, S^{T}_3\rangle&=&
\sqrt{\frac{I^T+1}{dim\,I^T}}\; 
\mid I_c=I^{T}+1/2;\;I^{T} \;I^{T}_3;\;\;S^{T}=I^{T} \; S^{T}_3 \rangle 
\nonumber\\ &+&\sqrt{\frac{I^T }{dim\,I^T}}\; 
\mid I_c=I^{T}-1/2;\;I^{T} \;I^{T}_3;\;\;S^{T}=I^{T}  \; S^{T}_3 \rangle
\nonumber\\
\mid K=1, I^{T}=S^{T}, I^{T}_3, S^{T}_3\rangle&=&
\sqrt{\frac{I^T }{dim\,I^T}}\; 
\mid I_c=I^{T}+1/2;\;I^{T} \;I^{T}_3;\;\;S^{T}=I^{T} \; S^{T}_3 \rangle 
\nonumber\\ &-&\sqrt{\frac{I^T+1 }{dim\,I^T}}\; 
\mid I_c=I^{T}-1/2;\;I^{T} \;I^{T}_3;\;\;S^{T}=I^{T} \; S^{T}_3 \rangle 
\end{eqnarray}
where $dim\;I\equiv 2I+1$.

 The matrix elements of 
 static operators between states of a  given spin-isospin tower  
      can be 
 expressed in terms of   matrix elements of 
 appropriate products of $SU(4)\times O(3)$ generators
 ($O(3)$ is generated by the orbital angular momentum operators). 
  An illustration of this claim is the familiar case of  
    matrix elements of the quadrupole moment operator between states
 of a given total angular momentum; they can be expressed in terms of 
 the matrix elements of the operator 
 $\{\hat{J}_i,\; \hat{J}_j\}-\frac{2}{3} \delta_{ij} \hat{J}^2$.

 At the level of one-body operators 
there is only one type of Hamiltonian that can give leading order spin-flavor symmetry
 breaking, namely, the Thomas precession   spin-orbit operator:  
\begin{equation}
H_{LS} \propto \;\hat{a}^{\dagger} \vec{L}.\vec{\sigma}\; \hat{a}~~~,
\end{equation}
where the proportionality factor is of  order $\Lambda_{{\rm QCD}}$. 
Here we assume that indices are
 contracted in the obvious way. This operator could be written 
 in the most general form (4), but for our purpose we only need its spin-flavor
 structure.
 Two-body operators of the same LS type can be constructed by
 multiplying the form 
given above by a bilinear  $\hat{a}^{\dagger} \hat{a}$ 
(which is the Baryon number operator
if no momentum is exchanged), properly generalized to the form given in  (6). 
These two-body operators  can be taken into account   by simply
 rescaling the matrix elements obtained with the one-body operator.

Symmetry breaking due to the interaction of the excited quark with the 
${\rm N_c}-1$
 quarks in the core starts   at the level of two-body operators. 
 The leading   operators are those 
that have coherent matrix elements  (${\cal{O}}({\rm N_c})$ enhancement) 
 between  states with $I^T<<{\rm N_c}/2$. The only coherent SU(4) generators  
  are the axial currents  
\begin{equation}
G^{ia}=\hat{a}^{\dagger}\sigma^i \otimes \tau^a \,\hat{a}~~~,
\end{equation}where $\sigma^i$ and $\tau^a$ are the spin and isospin 
Pauli matrices. 
 These operators naturally play a central role in the 
 study of 
  couplings between Baryons and pions$\cite {Sakita}$. 
Thus, the matrix elements of the leading operators of interest must 
include a factor of $G^{ia}$ acting on the core.  

There are four effective Hamiltonian structures  with factors of  $G^{ia}$   
we can write: 
\begin{eqnarray}
H_{T} &\propto& \frac{1}{{\rm N_c}}\; G^{ia} G_{ia}\nonumber\\
H_{1} &\propto & \frac{1}{{\rm N_c}} \;\hat{a}^\dagger L^i
\otimes \tau^a\,\hat{a}\;G_{ia} \nonumber\\
H_2  &\propto & \frac{i}{{\rm N_c}}\;  \hat{a}^\dagger [L^i,\, L^j] \otimes
\sigma_i 
\otimes \tau^a\,\hat{a}\;G_{ja} 
\nonumber\\
H_3  &\propto & \frac{1}{{\rm N_c}}\;  \hat{a}^\dagger \{L^i,\, L^j \} \otimes
\sigma_i 
\otimes \tau^a\,\hat{a}\;G_{ja} 
\end{eqnarray}
where the proportionality factors are of order $\Lambda_{{\rm QCD}}$.
As we show in a moment, the operator $H_T$ gives  mass 
contributions  
that   respect 
spin-flavor symmetry up to $1/{\rm N_c}$ terms,  
whereas the other operators   break the symmetry at leading order.

Higher body operators can be reduced to one- and two-body operators.
To show this, we   notice     that n-body operators will contain 
a total of (n-1) factors
of the coherent operators  $G_{ia}$ (cases where a factor of
the Baryon number operator appears are  clearly equivalent to 
 an (n-1) body operator, since this factor is equivalent to
  multiplying by a factor ${\rm N_c}$). Since the isospin piece
  associated with the matrix element between the excited quark states
  can   carry at most one isospin index $a$, products of  $G_{ia}$
  must have their isospin indices contracted in  such  a way that
  at most one such index is left.  Using the relations $\cite{Dashen2}$:
\begin{eqnarray}
G^{ia} G^{ja}=N_c^2\delta_{ij}\nonumber\\
G^{ia} G^{ib}=N_c^2\delta_{ab}\nonumber\\
\epsilon_{ijk}G^{ia} G^{jb}=N_c \epsilon_{abc} G^{kc}\nonumber\\
\epsilon_{abc}G^{ia} G^{jb}=N_c \epsilon_{ijk} G^{kc}\nonumber\\
\epsilon_{ijk}\epsilon_{abc}G^{ia} G^{jb}=2 \;N_c G^{kc},
\end{eqnarray}
it is obvious how to carry out the reduction of the
 higher body operators acting on the core. We see that at leading order
 only one-body operators acting on the core and
 proportional to the axial current or Baryon number
  are left. Thus, the one- and two-body operators listed before
  are the complete set we need  to discuss.

The matrix elements of $G^{ia}$ between core states were obtained by 
Dashen, Jenkins and Manohar$\cite{Dashen2}$: 
\begin{eqnarray}
&&\langle I_c^{\prime}=S_c^{\prime},\;I_{c3}^{\prime},
\;S_{c3}^{\prime}\mid G_{ia}\mid 
 I_c=S_c,\;I_{c3},\;S_{c3}\rangle
\nonumber\\
&=& {\rm N_c} \left[\frac{{\rm dim}\, I_c}{{\rm dim}\,I_c^{\prime}}\right]^{1/2}
\langle I_c\;I_{c3},\;1\;a\mid I_c^{\prime}\;I_{c3}^{\prime}\rangle\;
\langle I_c\;S_{c3},\;1\;i\mid I_c^{\prime}\;S_{c3}^{\prime}\rangle + {\cal O}(1)\nonumber\\
&& {\rm dim}\, j\equiv 2\,j+1~~~,
\end{eqnarray}
where the term of ${\cal O}(1)$ is actually proportional to the leading term.
 With this,    
the matrix elements of the   operators (12) between degenerate states
in the spin-flavor symmetry limit are straightforward to calculate. 
The explicit calculation in the case of $H_T$ gives:
\begin{eqnarray}
&&\langle I_c^{\prime};\;I^{T}\;I^{T}_3;\;\;S^{ T \prime} \;
 S^{T \prime}_3,\;\ell\;m^{\prime}
\mid H_T \mid
I_c;\;I^{T}\;I^{T}_3;\;\;S^{T} \; S^{T}_3,\;\ell\;m \rangle
\nonumber\\
&=&
\Lambda_{T}\; \delta_{S^{T \prime},S^{T}} \;\delta_{S^{T \prime}_3 ,S^{T}_3}\; \delta_{m,m^{\prime}}
(-1)^{1 -S^T-I^T} \; \left[{\rm dim} \,I_c\;{\rm dim}\, I_c^{\prime}\right]^{1/2} \nonumber\\
&\times & \left\{ \begin{array}{ccc} S^T&I_c& 1/2\\ 1 & 1/2 & I_c^{\prime}\end{array}\right\}
\left\{ \begin{array}{ccc} 
I_c&1& I_c^{\prime}\\ \frac{1}{2} & I^T & \frac{1}{2}\end{array}\right\}~~,
\end{eqnarray}
where $I_c,$ $I_c^{\prime}=I^T\pm 1/2$, $J^T$ is the total angular momentum of the Baryon, 
and $j$ is the total angular momentum of the excited quark. 

Using (15) and (9), we readily check that $H_T$ is   diagonal in the basis
of $K=0$ and $K=1$ states, and the eigenvalues are respectively       
 $(-1)^{2 S^T}/2$ and $(-1)^{2 S^T+1}/6$. 
 The contribution  of order ${\rm N_c}$   is spin-flavor independent
 up to order $1/{\rm N_c} $ corrections
and the same for K=0 and K=1 multiplets. This   contribution stems from
 $H_T$ acting on the core, and 
  can be  understood  
 by considering an infinitesimal SU(4) transformation on $H_T$. 
 The infinitesimal
transformation of zeroth order   generated by $X^{ia}=G^{ia}/{\rm N_c}$
produces an  energy  shift proportional to $[H_T,\; X^{ia}]$, which is
of order $1/{\rm N_c} $ as a consequence of the commutators 
$[G^{ia},\;G^{jb} ]$  having matrix elements of the order of the spin 
and isospin of the states. A similar  argument  implies that 
 the result (15) must respect  SU(4) symmetry.

The   matrix elements of the rest of the operators are lengthy but straightforward    
to calculate and read  as follows:
  
\begin{eqnarray}
&& \langle I_c^{\prime};\;I^T \; I^T_3; \; \ell, \; S^{T \prime}, \; J^T \; J^T_3 \mid H_{LS} \mid
I_c; \;I^T \; I^T_3; \; \ell, \; S^T, \; J^T \; J^T_3 \rangle \nonumber\\
 &=& \Lambda_{LS} \;\delta_{I_c, I_c^{\prime}}
 \left[ {\rm dim}\, S^T \; {\rm dim} \, S^{T \prime}\right]\nonumber\\
&\times& \!\!\sum_{j=\ell \pm 1/2} 
{\rm dim} \,j \;\left(j(j+1)-\ell (\ell+1)-3/4\right) 
\left\{ \begin{array}{ccc} I_c & 1/2 & S^T \\ \ell & J^T & j\end{array}\right\}
\left\{ \begin{array}{ccc} I^c & 1/2 & S^{T \prime}\\ \ell & J^T & j\end{array}\right\}
\nonumber\\ &&\nonumber\\
& & \langle I_c^{\prime};\;I^T \;I^T_3;\;\ell,\;S^{T \prime},\;J^T\;J^T_3\mid H_{1}\mid
I_c;\;I^T\;I^T_3;\;\ell,\;S^T,\;J^T\;J^T_3 \rangle \nonumber\\
&=&  \sqrt{\ell (\ell+1)}\; 
(-1)^{I^T-J^T+S^T-S^{T \prime}+I_c-I_c^{\prime}-\ell}
 \left[ {\rm dim}\, S^T\; {\rm dim}\, S^{T \prime}\; {\rm dim}\, I_c\;
  {\rm dim}\, I_c^{\prime}
\; {\rm dim}\, \ell\right]^{1/2}\nonumber\\
&\times& \left\{ \begin{array}{ccc}  I_c & 1 &  I_c^{\prime}\\  1/2 & I^T  & 1/2
\end{array}\right\}\;
\left\{ \begin{array}{ccc}  S^T& 1  & S^{T \prime}  \\
 I_c^{\prime}  & 1/2  & I_c  \end{array}\right\}\;
\left\{ \begin{array}{ccc}  \ell & 1 & \ell \\   S^{T \prime} & J^T  & 
S^T \end{array}\right\} \nonumber\\ & &\nonumber\\
&&\langle I_c^{\prime};\;I^T \;I^T_3;\;\ell,\;S^{T \prime},\;J^T\;J^T_3\mid H_{2}\mid
I_c;\;I^T\;I^T_3;\;\ell,\;S^T,\;J^T\;J^T_3 \rangle \nonumber\\
&=& 
\frac{3}{2} [{\rm dim}\,S^T\;{\rm dim}\, S^{T \prime}\;
{\rm dim}\, I_c\;{\rm dim}\, I_c^{\prime}]^{1/2} 
 \left\{ \begin{array}{ccc}  I_c & 1 &  I_c^{\prime}\\  1/2 & I^T  & 1/2
\end{array}\right\}\;
\sum_{j=\ell\pm 1/2}\;{\rm dim}\,j\;j(j+1)\nonumber\\
&\times&    (-1)^{I^T-S^{T\prime}}\;
\left\{ \begin{array}{ccc}  \frac{1}{2}  & \frac{1}{2}  & 1 \\ I_c  & I_c^{\prime} &  1
\end{array}\right\}\;
\left\{ \begin{array}{ccc}   I_c  & \frac{1}{2}  & S^T \\  \ell  &  J^T  & j 
\end{array}\right\}\;
\left\{ \begin{array}{ccc}   S^{T \prime} & \frac{1}{2}  & I_c  \\  j  & J^T   &  \ell
\end{array}\right\}\;-(I_c \leftrightarrow I_c^{\prime},S^T \leftrightarrow S^{T \prime})
\nonumber\\& &\nonumber\\
&&\langle I_c^{\prime};\;I^T \;I^T_3;\;\ell,\;S^{T \prime},\;J^T\;J^T_3\mid H_{3}\mid
I_c;\;I^T\;I^T_3;\;\ell,\;S^T,\;J^T\;J^T_3 \rangle \nonumber\\
&=&\sqrt{\frac{15}{2}}\;
(-1)^{\ell+J^T+I^T-\frac{1}{2}+I_c^{\prime}+S^{T \prime}+2 S^T}\;\nonumber\\ 
&\times&\sqrt{\frac{\ell (2 \ell -1)}{(\ell+1)(2\ell+3)}}
[{\rm dim}\,\ell\;{\rm dim}\,S^T\;{\rm dim}\, S^{T \prime}\;
{\rm dim}\, I_c\;{\rm dim}\, I_c^{\prime}]^{1/2}\nonumber\\
&\times&
\left\{ \begin{array}{ccc}   I_c^{\prime}    &  1   & I_c \\ 
\frac{1}{2}  & I^T &  \frac{1}{2}
\end{array}\right\}\;
\left\{ \begin{array}{ccc}   2  & \ell  & \ell \\ 
J^T  & S^{T \prime} & S^T
\end{array}\right\}\;
\left\{ \begin{array}{ccc}    S^{T \prime}  &  S^T  &  2 \\
 \frac{1}{2} & \frac{1}{2}  &  1 \\  I_c^{\prime}  & I_c   & 1 
\end{array}\right\} 
\end{eqnarray}
 
From the above we conclude that spin-flavor symmetry is broken at leading order only
in the case of Baryons with non-vanishing orbital angular momentum.

When $\ell>0$, the multiplicities of states are as follows:
in symmetric SU(4) representation, one state for each $\mid \ell-I^T\mid\leq
J^T\leq \ell+I^T$; in mixed symmetry representation, one state for
$J^T=\ell+I^T+1$ and for $J^T=min\{\mid \ell-I^T \pm 1\mid \}$, two for 
$J^T=\mid \ell-I^T \mid $, and three for the rest.

For illustration, we show the results for a few states
 in the case of $\ell=1$ excitations.  The states in the totally symmetric
 SU(4) representation turn out to be  affected  only by $H_3$.
 The (I,J) states (1/2,1/2) and (1/2,3/2) are not shifted, while
 (3/2,1/2) is shifted by $a_3/23$ and (3/2,3/2) is shifted by $-a_3/40$.
  Here, $a_i$ parameterizes  the energy shift associated   with  $H_i$.  
 The mixed symmetry states show a more complicated pattern of shifts. 
 This is due to the fact that the three effective Hamiltonians $H_{LS}$,
 $H_1$ and $H_3$ do contribute to the shifts, and that the  shown 
 (I,J) states have   multiplicities
   2 or 3. The texture of the mass matrices are the following:
 \begin{eqnarray}
 (1/2,1/2)&:&\nonumber\\  &&a_{LS}\;\left[ \begin{array}{cc} 
 -\frac{2}{3} & -\frac{1}{3}   \\  
  -\frac{1}{3}& -\frac{5}{6}\end{array}\right]
  +a_1\;\left[ \begin{array}{cc} \frac{2}{3} & -\frac{1}{3}  \\  
 -\frac{1}{3}  & \frac{5}{6}\end{array}\right]
 +a_3\;\left[ \begin{array}{cc} 0 & -\frac{1}{24\sqrt{2}} \\  
 -\frac{1}{24\sqrt{2}} & -\frac{1}{12}\end{array}\right]\nonumber\\&&\nonumber\\ 
 (1/2,3/2)&:&\nonumber\\ 
  &&a_{LS}\;\left[ \begin{array}{cc} 
 \frac{1}{3} & -\frac{\sqrt{5}}{6}   \\  
  -\frac{\sqrt{5}}{6} & -\frac{1}{3}\end{array}\right]
  +a_1\;\left[ \begin{array}{cc} 
 - \frac{1}{3} &  -\frac{\sqrt{5}}{6} \\  
   -\frac{\sqrt{5}}{6} & \frac{1}{3} \end{array}\right]
 +a_3\;\left[ \begin{array}{cc} 0 & \frac{1}{48\sqrt{5}} \\  
  \frac{1}{48\sqrt{5}} & \frac{1}{15}\end{array}\right]\nonumber\\&&\nonumber\\ 
  (3/2,1/2)&:& \\  &&
 a_{LS}\;\left[ \begin{array}{cc} 
 -\frac{1}{3} &  \frac{\sqrt{5}}{6}   \\  
  \frac{\sqrt{5}}{6} &  \frac{1}{3}\end{array}\right]
  +a_1\;\left[ \begin{array}{cc} 
   \frac{1}{3} & \frac{\sqrt{5}}{6}   \\  
  \frac{\sqrt{5}}{6}  & -\frac{1}{3} \end{array}\right]
 +a_3\;\left[ \begin{array}{cc} 
  \frac{17}{480}   & -\frac{5}{96}  \\  
   -\frac{5}{96} & 0\end{array}\right]\nonumber\\&&\nonumber\\ 
 (3/2,3/2)&:&\nonumber\\  &&\nonumber\\ 
 a_{LS}&&\left[ \begin{array}{ccc} 
- \frac{2}{15}  & \frac{5}{6 \sqrt{2}}  &  -\frac{3 \sqrt{3}}{10 \sqrt{2}} \\  
    \frac{5}{6 \sqrt{2}} &- \frac{1}{6}  & 0  \\  
  -\frac{3 \sqrt{3}}{10 \sqrt{2}}   &  0  & -\frac{7}{10} \end{array}\right]
  +a_1\;\left[ \begin{array}{ccc} 
   \frac{2}{15} &  \frac{5}{6\sqrt{2}}   &  -\frac{3 \sqrt{3}}{10\sqrt{2}}  \\  
   \frac{5}{6\sqrt{2}}    & \frac{1}{6}  & 0  \\  
     -\frac{3 \sqrt{3}}{10\sqrt{2}} & 0  & \frac{7}{10}  \end{array}\right]
 +a_3\;\left[ \begin{array}{ccc} 
    - \frac{17}{600}& \frac{1}{96\sqrt{2}}   & - \frac{21\sqrt{3}}{800\sqrt{2}}  \\   
    \frac{1}{96\sqrt{2}}  & 0  & 0  \\  
     - \frac{21\sqrt{3}}{800\sqrt{2}} &  0   
     & -\frac{7}{150}   \end{array}\right] \nonumber
  \end{eqnarray}
 Quite in general, irrespective of the value of $\ell$, the symmetric states are
 affected only by $H_3$, while the mixed symmetry states turn out to be
  insensitive to $H_2$. Thus, in first order of perturbation theory only 
  three operators are relevant.
 
So far we have only considered  matrix elements between states
  degenerate in the   spin-flavor symmetry limit. Beyond first order of 
  perturbation theory 
  the symmetry breaking  
    produce leading order   mixings  between   states belonging to
  different spin-flavor multiplets. In this case, other operators 
  besides the ones considered  so far become relevant. In particular,
  we expect that the $\ell=0$ states will experience spin-flavor breaking
  through the mixing with states in other towers.  The problem of
  going beyond first order perturbations will be addressed elsewhere.
  
  Finally,
concerning the origin of the
effective interactions discussed in this section, except for 
$H_{LS}$, the other two interactions do not emerge directly from QCD by
performing the $1/M_{\rm quark}$ expansion  and must, therefore, 
crucially   depend on the
structure of the  constituent light quark. 
While the spin-orbit piece
 is   included in   analyses  of Baryon spectroscopy $\cite{Isgur}$,  
$H_{T}$ has been realized by a pion exchange model $\cite {Riska}$
 which gives a-priori big relevance to this interaction, 
 and, the operators  $H_{i},~(i=1,2,3)$
 have not been considered as far as we know.

\section{Conclusions}

The spin-flavor symmetry breaking level shifts of non-strange excited 
Baryons have been discussed in the large
${\rm N_c}$ limit to leading order in the $1/{\rm N_c}$ expansion 
and first order in perturbation theory.
 In this approximation,    only 
Baryons with non-vanishing orbital angular momentum do experience
leading order spin-flavor breaking in their levels. 
  It was also shown that it is necessary to include only up to two-body operators
as higher body operators can  be reduced. Similar results will hold for three flavors.
  Our  study  is  of theoretical interest and we expect
   that they can be useful
in understanding other large ${\rm N_c}$ studies of QCD, for instance,
 through   lattice   simulations. 
 Finally, it would be interesting to go beyond first order of
 perturbation theory and also  to know  how important are the effects
of the   effective interactions discussed here
for the case of real Baryons.

\section{Acknowledgements}
I thank   M. Burkardt, S. Capstick, T. D. Cohen, O. W. Greenberg, 
N. Isgur, R. Lebed, 
K. M. Maung, W. Roberts, R. Schiavilla, J. Tjon and J. D. Walecka for useful 
remarks and discussions.
This work has been supported in part by NSF grant HRD-9633750.

\def\prd#1#2#3{{\it Phys. ~Rev. ~}{\bf D#1} (19#2) #3}
\def\prc#1#2#3{{\it Phys. ~Rev. ~}{\bf C#1} (19#2) #3}
\def\plb#1#2#3{{\it Phys. ~Lett. ~}{\bf B#1} (19#2) #3}
\def\npb#1#2#3{{\it Nucl. ~Phys. ~}{\bf B#1} (19#2) #3}
\def\npa#1#2#3{{\it Nucl. ~Phys. ~}{\bf A#1} (19#2) #3}
\def\prl#1#2#3{{\it Phys. ~Rev. ~Lett. ~}{\bf #1} (19#2) #3}


\begin{thebibliography}{99}
\bibitem {t'Hooft} G.  t'Hooft, Nucl. Phys. B72 (1974) 461. 
\bibitem {1+1} G.  t'Hooft, Nucl. Phys. B75 (1974) 461.
\bibitem {Callan} C. G. Callan, N. Coote and D. Gross, Phys. Rev. D13 (1976) 1649.\\
M. B. Einhorn, Phys. Rev. D14 (1976) 3451.
 \bibitem{Coleman} S. Coleman,  in \lq\lq Pointlike Structures Inside
 and Outside  Hadrons".  Proceedings 
   edited by A. Zichichi. Plenum Press, (1982) 11.
\bibitem{Witten1} E. Witten, Nucl. Phys. B160 (1979) 57.
\bibitem{Skyrme} T. H. Skyrme, Proc. R. Soc. A260 (1961) 127.
\bibitem{Witten2} E. Witten, Nucl. Phys. B223 (1983) 433.
\bibitem{Manohar1} A. V. Manohar, Nucl. Phys. B248 (1984) 19.
\bibitem{Sakita} J. L. Gervais and B. Sakita, Phys. Rev. Lett. 52 (1984) 87, and Phys. Rev. D30 (1984) 1795.\\
K. Bardakci, Nucl. Phys. B243 (1984) 197.
\bibitem{Dashen1} R. Dashen and A. V. Manohar, Phys. Lett. B315 (1993) 425, 438.
\bibitem{Jenkins} E. Jenkins, Phys. Lett. B315 (1993) 431, 441, 447.\\
 E. Jenkins and R. F. Lebed, Phys.Rev.D52 (1995) 282.
\bibitem{Lebed} M. A. Luty, J. March-Russell and M. White, Phys. Rev. D51 (1995) 2332.\\
E. Jenkins and A. V. Manohar, Phys. Lett. B335 (1994) 452. \\
 R. F. Lebed, Phys.Rev.D51 (1995) 5039.
\bibitem{Carone} Ch. D. Carone, H. Georgi, L. Kaplan and D. Morin, Phys. Rev. D50 (1994) 5793.
\bibitem{Masperi} F. D. Mazzitelli and L. Masperi, Phys. Rev. D35 (1987) 368.\\
  W. Broniowski, Nucl.Phys. A580 (1994) 429.\\
A. Wirzba, M. Kirchbach and D.O. Riska, J. Phys. G20 (1994) 1583. 
\bibitem{Mattis} M. P. Mattis and M. Mukherjee, Phys. Rev. Lett. 61 (1988) 1344.
\bibitem{Cook}T. Cook, C. J. Goebel and B. Sakita, J. Math. Phys. 8 (1967) 708.
\bibitem{Dashen2} R. Dashen, E. Jenkins and A. V. Manohar, Phys. Rev. D49 (1994) 4713, and Phys. Rev. D51 (1995) 3697.
\bibitem{Luty}  M. A. Luty and J. March-Russell, Nucl. Phys. B426 (1994) 71.
\bibitem{Isgur} N. Isgur and G. Karl, Phys. Rev. D18 (1978) 4187, and  Phys. Rev. D19 (1979) 2653.\\
S. Capstick and N. Isgur, Phys. Rev. D34 (1986) 2809.
\bibitem{Riska} L. Ya. Glozman and D. O. Riska, Phys.Rept. 268 (1996) 263, and references therein.

\end{thebibliography}
\end{document}